\documentclass[prb,twocolumn,amsmath,amssymb,floatfix,superscriptaddress]{revtex4-2} 
\usepackage{graphicx}
\usepackage{epstopdf}  
\usepackage{bm}
\usepackage{amsmath,amssymb}
\usepackage{dcolumn}
\usepackage{bm}
\usepackage{epsfig}
\usepackage{longtable}
\usepackage{subfigure}
\usepackage{multirow}
\usepackage{color}

\newcommand{\bfq}{ {\bf q}} 
\newcommand{\bfk}{ {\bf k}} 

\begin{document}

\title{{\rm\small\hfill }\\
   Phase Stability of the Argon Crystal: A First-Principles Study Based on Random Phase Approximation plus Renormalized Single Excitation Corrections }

\author{Sixian Yang}
\affiliation{Key Laboratory of Quantum Information, University of Science and Technology of China, Hefei, 230026, China}
\author{Xinguo Ren}
\email{renxg@iphy.ac.cn}
\affiliation{Beijing National Laboratory for Condensed Matter Physics, Institute of Physics, Chinese Academy of Sciences, Beijing 100190, China
}


\begin{abstract}
The energy differences between the face-centered cubic (FCC) and hexagonal closed packed (HCP) structures of the argon (Ar) crystal are studied using the first-principles
electronic-structure
approach at the level of random phase approximation (RPA) plus renormalized single exitation (rSE) correction.  By treating both structures at equal footing (i.e., 
employing the same computational supercell and $\bfk$ grid sampling), our RPA+rSE calculations show that, at zero temperature, the FCC structure is lower in energy than the HCP
structure over a wide pressure range. The influence of zero-point energy (ZPE) is also studied and and it is found that ZPE only plays a secondary role in determining
the relative stability of the two structures, whereas the electron correlation effect dominates. We further examine the equation of states in the high pressure regime,
and our RPA+rSE results, complemented with phonon contributions, show excellent agreement with available experimental data. Finally, by computing the Gibbs free energies 
for both the FCC and HCP at different temperatures, we are able to generate a $T$-$P$ phase diagram for the Ar crystal, disclosing the pressure-temperature range 
for each phase. Our calculations reveal that the FCC phase has a slightly larger entropy and volume than HCP phase at the same temperature and pressure condition.
\end{abstract}

\maketitle
\section{Introduction}
An essential goal of computational physical science is to reliably predict the relevant polymorphs and their 
relative stability, and consequently the thermodynamic behavior of a substance, with only knowledge of its chemical composition. 
This challenge largely persists even today after it was noted over 30 years ago \cite{Maddox:1988}. Nevertheless, considerable progress has been
made towards this goal, thanks to the development of various structure prediction algorithms
\cite{Wille:1987,Wales/Doye:1997,Deaven/Ho:1995,Glass/Oganov/Hansen:2006,WangYC/etal:2010} 
as well as the continuing methodology and accuracy advances of first-principles calculations  \cite{Burke:2012,Jones:2015,Heyd/Scuseria/Ernzerhof:2003,Sun/etal:2016,Lejaeghere/etal:2016,Marzari/etal:2021}. 
Nowadays, it is not uncommon for the mainstream crystal structure prediction (CSP) methods to predict the relevant known and unknown 
polymorphic forms of a material, as well as their energy separations, 
starting with a pre-given chemical composition. However, the output of such predictions, in particular the energy ranking of different
polymorphs will necessarily depend on the underlying electronic structure methods for computing the potential energy surface. 
For instance, most CSP algorithms can predict the existence of diamond and graphene allotropes for carbon, but 
cannot tell graphene is thermodynamically more stable at ambient conditions, if conventional local/semilocal density functional 
approximations  (DFAs) are employed as the electronic structure solver. Similar issues occurs if one quests for whether
the face-centered cubic (FCC) or hexagonal close packed (HCP) crystal structure is preferred for rare gas crystals under
ambient pressure \cite{Klein1976Rare,PhysRevLett.67.3263}. Thus, in the long-term endeavor aiming at quantitatively and faithfully predict the 
thermodynamical behavior of real materials without invoking empirical inputs, further improving the accuracy and reliability of 
the underlying electronic structure methods that can be conveniently used together with the CSP algorithms is a must.

In fact, the above-noted phase stability of rare-gas crystals has posed a significant challenge to first principles electronic structure methods.
In particular, it is theoretically highly nontrivial to describe the delicate energy differences between the FCC and HCP phases at
zero temperature, for which a sub-meV accuracy is mandatory. Without a solid base at 0 K, it is then highly unlikely 
that one can convincingly explain why rare gas atoms prefer to crystallize in the FCC phase instead of the HCP phase under ambient pressures, 
let alone the entire
phase diagram of the rare-gas systems under varying thermodynamic conditions. Within the realm of first principles electronic structure theories,
conventional local and semilocal DFAs \cite{Kohn/Sham:1965,Perdew/Burke/Ernzerhof:1996} are unable to reliably describe the cohesive properties 
of rare-gas systems due to their inadequate treatment of long-range
van der Waals (vdW) interactions. Complementing semi-local functionals with semiempirical vdW corrections
\cite{Becke/Johnson:2007,Grimme/etal:2010,Tkatchenko/Scheffler:2009,Tkatchenko/etal:2012} or non-local vdW density functionals
\cite{Dion/etal:2004,Klimes/Bowler/Michaelides:2010} can 
significantly improve their usefulness in describing vdW-bonded systems, but it is questionable whether these
pragmatic vdW-inclusive schemes can provide the needed (sub-meV) accuracy in order to describe the phase stability of
rare-gas crystals. Beyond these approximations one may either go to the fifth-rung functionals
\cite{Perdew/Schmidt:2001} (e.g., the random phase approximation (RPA) 
\cite{Bohm/Pines:1953,Langreth/Perdew:1977,Gunnarsson/Lundqvist:1976,Eshuis/Bates/Furche:2012,Ren/etal:2012b} and 
beyond \cite{Ren/etal:2012,Ren/etal:2013}) within the realm of the density functional theory (DFT)
where non-local electron correlation effects are seamlessly incorporated, or simply go to wave-function based
approaches, like the second-order M{\o}ller-Plesset perturbation theory (MP2) \cite{Moller/Plesset:1934} and the coupled cluster (CC) theory \cite{Bartlett/Musial:2007}. However, MP2, which contains only
two-body (here ``body" means atom) correlations, is considered not 
sufficiently accurate to describe the cohesive properties of rare-gas crystals \cite{Hermann2009Complete,CasassaA,PhysRevB.82.205111}.

In the connection,  the CC theory 
truncated at the level of single, double, and perturbative triple excitations (CCSD(T)) \cite{Head-Gordon/Pople:1988}
(known as ``gold standard" in computational chemistry) is of great interest and has been applied 
to study the cohesive properties of rare-gas systems \cite{Hermann2009Complete,CasassaA,PhysRevB.82.205111,PhysRevLett.79.1301,PhysRevB.62.5482,PhysRevB.60.7905,Schwerdtfeger2016Towards,PhysRevB.73.064112,PhysRevB.95.214116}. 
However, in these studies, instead of directly applying CCSD(T) 
 to treat infinite periodic rare-gas crystals,  one utilizes the method only to generate the short-range part of 
the potential energy landscapes of trimers and tetramers. The entire cohesive energies are obtained by summing up two-body, three-body,
and four-body contributions, with each of them being calculated by summing over sufficiently large atomic clusters in real space.
The interactions between atomic pairs, as well as the long-range part of the interactions within trimers and tetramers are being described by empirical model potentials.
Thus this is a rather sophisticated approach that combines many ingredients into a framework, and necessarily depends on many parameters.
Despite the remarkable accuracy this approach delivers and its successful explanation of the preference to the FCC phase of rare-gas crystals \cite{Schwerdtfeger2016Towards}, it is rather tedious to use for non-experts and barely suitable for routine applications, in particular
in the context of high-throughput computations.

In this work, we present a detailed study of the performance of RPA and beyond for computing 
the cohesive energies of the Ar crystal, with special emphasis on 
the energy differences between the FCC and HCP phases. Here RPA refers to a theoretical approach for 
computing the ground-state total
energy of many-electron systems, in a way that the exchange energy is evaluated exactly and
the electron correlation energy is treated at the direct RPA level.
The exchange-correlation (XC) energy of the RPA method can be viewed as an orbital-dependent energy functional 
derived within the framework of adiabatic-connection fluctuation-dissipation theorem of DFT \cite{Langreth/Perdew:1977,Gunnarsson/Lundqvist:1976,Dobson:1994}, but
it is also intimately connected to the CC theory \cite{Scuseria/Henderson/Sorensen:2008}. One particularly appealing aspect of RPA is that it 
provides a balanced description of
all bonding characteristics, including a seamless incorporation of vdW interactions. Early benchmark studies
indicated that RPA (and beyond) is capable of describing delicate energy differences \cite{Ren/etal:2012b}. 
This is corroborated by more recent studies that revealed that RPA yields the correct energy ranking of
different polymorphs of several types of materials, including carbon allotropes \cite{Lebegue/etal:2010}, FeS$_2$ \cite{Zhang/Cui/Zhang:2018}, BN \cite{Cazorla/Gould:2019}, as well as a few others \cite{Sengupta/Bates/Ruzsinszky:2018}. 

Despite its success, the standard RPA scheme shows an overall tendency of underestimating the binding energies of molecules and solids
\cite{Furche:2001,Harl/Schimka/Kresse:2010}, and this underestimation
can be largely alleviated by adding a correction term arising from single excitations \cite{Ren/etal:2011}. In particular, the renormalzied single excitation (rSE) 
correction has been shown to appreciably improve the accuracy of the standard RPA for describing the binding strengths
of molecules and solids \cite{Paier/etal:2012,Ren/etal:2013,Klimes/etal:2015}. In this work, we apply the RPA+rSE method \cite{Ren/etal:2013} to study the
energy difference between the FCC and HCP structures of the Ar crystal. Here, instead of resorting to a tedious cluster-based approach \cite{PhysRevLett.79.1301,PhysRevB.62.5482,PhysRevB.60.7905,Schwerdtfeger2016Towards,PhysRevB.73.064112,PhysRevB.95.214116}, we directly treat the crystal
as a periodic system in terms of $\bfk$ point sampling. We find that, in order to capture the delicate energy differences between the FCC and HCP structures, 
it is important to treat the two structures at an equal footing, i.e., setting up a computational cell that is the same for the FCC and HCP crystal structures and 
performing calculations with the $\bfk$ point setting. Test calculations show that using their primitive unit cells 
of these two structures (and hence different $\bfk$ point setting) will not provide sufficient numerical precision. With the above-noted computational strategy, and by carefully
converging all the computational parameters, we found that the RPA+rSE cohesive energy for the FCC phase is lower by around 3.6 $\mu$Ha ($\sim$ 0.1 meV) than that of
the HCP phase at zero temperature and ambient pressure. The zero-point energy (ZPE), which was previously suggested to be vital for stabilizing the FCC phase  \cite{PhysRevB.62.5482,PhysRevB.60.7905,Schwerdtfeger2016Towards},
is found to only play a secondary role here.
Furthermore, by including the contributions of phonon energies, we calculate the Helmholtz free energy and derive the pressure-volume ($P$-$V$) curves at room temperatures.
Our results show excellent agreement with available experimental data in the high-pressure regime, which was previously considered to be a challenging
problem for the Ar crystal \cite{PhysRevB.95.214116}. Finally, by computing the Gibbs free energy for both phases, we are able
to determine the phase boundary of the FCC and HCP phases in the temperature-pressure ($T$-$P$) phase diagrams. The behavior of the phase boundary line can be understood
from the perspective that the FCC phase has a slightly large entropy and volume that the HCP phase at the same temperature and pressure condition.

Our investigation suggests that the combination of advanced 
electronic structure calculations at the level of RPA and beyond with reliable estimations of phonon energies is a powerful approach to 
study the thermodynamic properties of condensed matter systems. Since the implementation of such approach has been available in several mainstream computer code
and the barrier of employing such an approach for routine application is not high, we expect this approach will become highly useful for treating challenging
problem when unprecedented accuracy is required.

\section{\label{sec:results}Results and Discussions}
\begin{figure}[t]
\centering
\includegraphics[width=0.45\textwidth,clip]{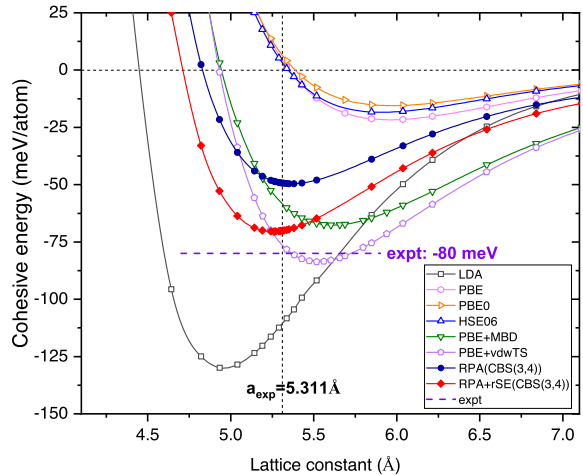}
\caption{\label{Fig:Cohesive} Cohesive energies of the FCC Ar crystal determined by different functionals. All calculations are done using
the FHI-aims code package \cite{Blum/etal:2009,Ren/etal:2012} and NAO-VCC-$n$Z basis sets \cite{IgorZhang/etal:2013}. The RPA and RPA+rSE results
are obtained by extrapolating the NAO-VCC-3Z and 4Z results to the complete basis set (CBS) limit, and  all other results are obtained using the NAO-VCC-4Z basis set. 
A primitive unit cell and $10\times 10 \times 10$ $\bfk$ grid sampling
are used in all calculations.
ZPE contribution is not included here. Experimental values are taken from Ref.~\cite{schwalbe1977thermodynamic}.}
\end{figure}

\textbf{Cohesive energy curves for the Ar crystal.}
We start by first looking at the performance of RPA+rSE for describing the cohesive property of the Ar crystal. In Fig.~\ref{Fig:Cohesive} 
we present the cohesive energies as a function of the lattice constant for the FCC Ar crystal, as determined by various
functionals, including  RPA@PBE and (RPA+rSE)@PBE.  Here ``@PBE" denotes that the calculations
are done on top of the reference wavefunctions generated by a preceding KS calculation based on the generalized gradient approximation (GGA) of Perdew, Burke, and Ernzerhof \cite{Perdew/Burke/Ernzerhof:1996}. 
For comparison, results obtained using other functionals, including local-density approximation (LDA), GGA-PBE itself, the global hybrid functional 
PBE0 \cite{Perdew/Ernzerhof/Burke:1996}, the Heyd-Scuseria-Ernzerhof (HSE) 
screened hybrid functional \cite{Heyd/Scuseria/Ernzerhof:2003},
as well as PBE complemented with the Tkatchenko-Scheffler (TS) vdW correction  (vdW(TS)) \cite{Tkatchenko/Scheffler:2009} and with the more sophisticated many-body dispersion (MBD) \cite{Tkatchenko/etal:2012}, are also presented. 
In addition, the experimental equilibrium lattice constant and cohesive energy are indicated by dashed lines in Fig.~\ref{Fig:Cohesive}. 
All calculations are done using the FHI-aims code package \cite{Blum/etal:2009,Ren/etal:2012}, based on an all-electron, numeric atom-centered 
orbital (NAO) basis set framework. The correlation-consistent NAO-VCC-$n$Z basis sets \cite{IgorZhang/etal:2013} are used in the calculations.
Since correlated methods like RPA and beyond have different basis set dependence behavior compared to
the conventional functionals, the presented RPA and RPA+rSE results are those extrapolated to the complete basis set (CBS) limit, 
whereas the results of other functionals are obtained with NAO-VCC-4Z basis set.
Further computational details of these calculations 
are given in Sec.~\ref{sec:method}. Figure~\ref{Fig:Cohesive} clearly shows that
a quantitatively accurate description of the cohesive properties of the Ar crystal is highly challenging for first-principles approaches. 
While LDA displays a pronounced artificial overbinding, GGA-PBE shows the opposite behavior. Going from PBE to the hybrid
functionals does not improve the situation, due to the fact that long-range vdW interactions are missing in both types of functionals.
Complementing the PBE functional with semi-empirical vdW corrections significantly improves the cohesive energies, as shown by
by PBE+vdW(TS) and PBE+MBD curves in Fig.~\ref{Fig:Cohesive},  but the equilibrium
lattice constant is still appreciably overestimated.  In case of RPA-based approaches, RPA@PBE performs significantly better than
LDA, GGA, and hybrid functionals, and the obtained equilibrium lattice constant ($5.365$ \AA) is in excellent agreement with the experimental value ($5.311$ \AA). 
However, similar to the Ar dimer (Ar$_2$) case \cite{Ren/etal:2011}, RPA@PBE underestimates the cohesive energies. Finally, when adding the rSE correction, this underbinding
is largely alleviated, and the (RPA+rSE)@PBE cohesive energy curve shows a rather satisfactory agreement with the experimental data, despite that
a slight underestimation of the cohesive energy is still noticeable. It should
be noted that RPA \cite{Harl/Kresse:2008} and RPA+rSE \cite{Klimes/etal:2015}  have been previously applied to the FCC phase of the Ar crystal, 
using an independent implementation based on projector augmented wave (PAW) method and plane-wave basis set \cite{Kresse/Furthmuller:1996a}. 
In their paper \cite{Klimes/etal:2015}, 
an even better agreement of the RPA+rSE cohesive energy with the experimental result was reported. We expect such a slight difference arising from the different numerical implementations does not affect the discussions below. 

\begin{figure}[t]
\centering
\includegraphics[width=0.45\textwidth,clip]{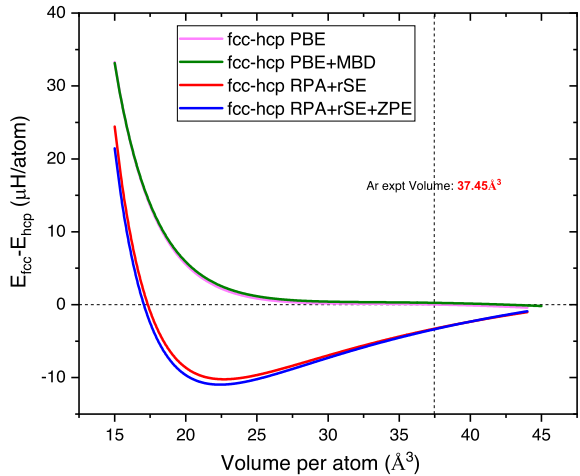}
\caption{\label{Fig:fcc-hcp_deltaE} Energy differences between the FCC and HCP structures of the Ar crystal at zero temperature as 
a function of the volume (per atom). Presented 
results are calculated using PBE, PBE+MBD, (RPA+rSE)@PBE approaches. Also shown are the (RPA+rSE)@PBE cohesive energy complemented with ZPE at the PBE level, denoted as ``RPA+rSE+ZPE". Results from RPA+rSE and ZPE@PBE are extrapolated to CBS(3,4), while those from other functionals are obtained using NAO-VCC-4Z basis set. The presented curves are obtained by subtracting the corresponding cohesive energy curves of the FCC and HCP structures,
which themselves are obtained by a second-order Birch-Murnaghan fitting of the original computed data.}
\end{figure}

\textbf{Energy differences between FCC and HCP structures.}
Encouraged by the excellent performance of RPA+rSE for describing the cohesive properties of the Ar crystal, we set out to 
examine the energy difference of the FCC and HCP structures of Ar using this approach. Here we stress that it is rather challenging to numerically capture
the delicate energy differences of the two structures, due to the fact that they have different primitive unit cells, 
and a finite $\bfk$-point sampling based on the primitive cells leads to an unbalanced description of the two structures.
To deal with this problem, we set up a $1\times 1 \times  6$ supercell for both structures, whereby the $z$ axis is chosen along the $[111]$ direction of
the FCC structure and the $c$ direction of the HCP structure, respectively. This results in a supercell with $ABCABC$ 
stacking for the FCC structure and with $ABABAB$ stacking
for the HCP structure. Then by utilizing the same setting for $\bfk$-point mesh ($12\times 12 \times 2$ in the present case), one achieves a numerical description of 
both structures on an equal footing.   
As a consequence, the target energy differences of the two structures are least prone to possible numerical errors arising from the under-convergence of computational parameters. 
In Fig.~\ref{Fig:fcc-hcp_deltaE}, we plot the energy differences of the cohesive energies of the FCC and HCP structures obtained by PBE, PBE+MBD, and (RPA+rSE)@PBE 
as a function of the volume. It can be seen that, most strikingly, the (RPA+rSE)@PBE yields a lower energy for the FCC phase than the HCP phase over a wide volume range. The energy preference to the FCC phase is approximately 3.5 $\mu$Ha at the ambient pressure, and reaches its maximum ($\sim$ 10 $\mu$Ha) at a compressed volume of approximately 22 \AA$^3$ per atom. The energy difference is reduced when the system is further compressed, 
until the HCP phase is favored at even higher 
pressures (smaller volumes). In contrast to the RPA+rSE results, the PBE and PBE+MBD yield almost identical cohesive energies for the two phases 
over a large volume range, and only at very high pressures
the HCP strutur becomes clearly favored. Furthermore, comparing the PBE and PBE+MBD results reveals that the long-range vdW interactions 
do not play a role
in discerning the energy difference of the two phases, although they do contribute significantly to the cohesive energies of each
individual phase. Last but not least, the comparison of RPA+rSE and PBE+MBD results provides additional insights. 
It should be noted that MBD captures the many-body coupling among the atoms at the RPA level, with each individual atom treated as
a harmonic oscillator\cite{Tkatchenko/etal:2012}. The different behaviors of PBE+MBD and (RPA+rSE)@PBE in Fig.~\ref{Fig:fcc-hcp_deltaE} suggests that 
it is really the many-body correlations at the electronic level that makes a qualitative difference here. 

In addition to the electronic energies, it has been pointed out in the past that the ZPE plays an
essential role in stabilizing the FCC phase over the HCP phase \cite{PhysRevB.62.5482,PhysRevB.60.7905,Schwerdtfeger2016Towards}. To check this point, 
we further add the differences of the ZPE contributions of the two phases to the RPA+rSE results, and plot the resultant RPA+rSE+ZPE results in Fig.~\ref{Fig:fcc-hcp_deltaE} (blue curve). 
It can be seen that the ZPE contribution indeed favors the FCC phase, shifting the entire $E_\text{FCC} - E_\text{HCP}$ curve further down.
However, its effect is much smaller in magnitude compared to the electron correlation effect described by RPA+rSE. It should be noted that in Fig.~\ref{Fig:fcc-hcp_deltaE}
the ZPE contributions are calculated using the PBE functional within the harmonic approximation. In principle, it would be ideal if the phonon 
frequencies were evaluated by the RPA+rSE approach. Unfortunately, at present
our implementation does not yet allow one to compute the phonon spectra according to the RPA+rSE potential energy surface, and evaluate the ZPE contributions at
the RPA+rSE level. However, as shown in Fig.~S1 of the supporting material (SM)
, we have checked the ZPEs using LDA, PBE and PBE+MBD functionals 
(among which the cohesive behavior of RPA+rSE sits in between), and confirmed that the ZPE contribution to the energy difference does not change appreciably with respect to the underlying functionals.
We thus expect that the scenario illustrated in in Fig.~\ref{Fig:fcc-hcp_deltaE} does not change if RPA+rSE were used to evaluate the ZPEs.
The anharmonic correction to the ZPE has also been considered in the literature \cite{Schwerdtfeger2016Towards}, but its effect is rather small
and neglected here.

\textbf{The $P$-$V$ curve in the high-pressure regime.}
Besides the mystery regarding the preferred crystal structure at ambient pressure, the rare-gas systems in the high-pressure regime also
exhibit fascinating phenomena. These include the pressure-induced structural phase transition from FCC to HCP (or to other structures at 
even higher pressures),  the change of electronic structure, e.g., the narrowing of the band gap and/or the change of conductance \citep{PhysRevLett.62.669}, as well as the rearrangement of electronic states \citep{PhysRevLett.95.257801}, etc. Even superconductivity
has been predicted to exist at very high pressures \citep{PhysRevB.91.064512}. 
Considerable efforts have been devoted to understanding the physical properties of rare gas systems under high pressure \citep{PhysRevB.95.214116,PhysRevLett.95.257801,PhysRevB.52.15165,PhysRevLett.88.075504,PhysRevLett.96.035504}.
In particular, Schwerdtfeger and coworkers \cite{Schwerdtfeger/Hermann:2009,PhysRevB.95.214116} derived the volume and temperature dependence of the the pressure --  $P(V,T)$
-- for both Ne and Ar crystals from the Helmholtz free energy, which consists of a static term determined from many-body expansion based on
the CC theory and a dynamical term determined from lattice vibrations. While excellent agreement with experiment was found for Ne up to
200 GPa \cite{Schwerdtfeger/Hermann:2009}, a comparable level of agreement was only achieved up to 20 GPa for Ar at room
temperature \cite{PhysRevB.95.214116}.
In the pressure range of
20-100 GPa, appreciable discrepancy between calculated results and experiment was observed. As such, a theoretical description of the equation
of state at high pressure reaching the experimental accuracy was considered to be a significant challenge for the  Ar crystal \cite{PhysRevB.95.214116}.

\begin{figure}[t]
\centering
\includegraphics[width=0.45\textwidth,clip]{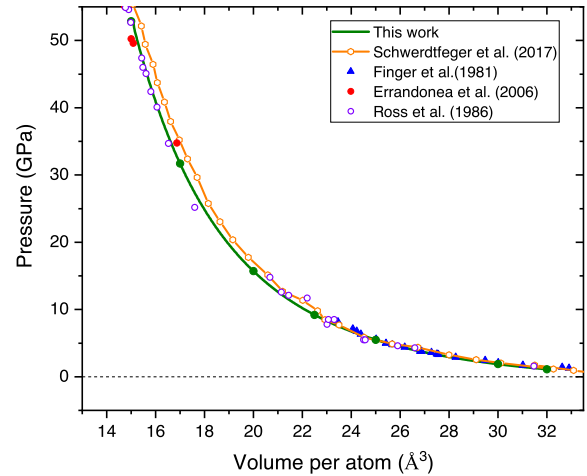}
\caption{\label{Fig:P-V_diagram} Calculated $P$-$V$ diagram  at 300 K for the FCC phase of the Ar crystal. Calculations are done using the 6-atom supercell 
(as Fig.~\ref{Fig:fcc-hcp_deltaE}) and the results are extrapolated to the CBS limit. We choose RPA+rSE accomplished with phonon free energy at 300K in our work here(Green line). Theoretical results of Ref.~\cite{PhysRevB.95.214116} 
obtained using many-body expansion and the experimental data taken from Ref.~\cite{FingerStructure,ross1986equation,Errandonea2006Structural} are included
for comparison.}
\end{figure}

To check how RPA+rSE performs in the high-pressure regime, we here also determine the equation of state $P(V,T)$ from the Helmholtz free
energy via $P(V,T)=-{\partial F(V,T)/\partial V}|_T$, whereby the expression of the Helmholtz free energy is presented in Sec.~\ref{sec:method}
(Eq.~\ref{Eq:QHA}). In brief,  the static part of the  Helmholtz energy in our calculations is given by RPA+rSE total energy, whereas
the ZPE and lattice thermodynamic vibration contributions are determined from the phonon spectrum based on PBE calculations. 
The possible influence of the underlying XC functional for
evaluating the phonon spectra on the $P$-$V$ curve is examined in Fig. S2 of the SM, whereby one can see again that varying the
functional has negligible impact. In Fig.~\ref{Fig:P-V_diagram}, our calculated $P(V,T)$ results 
as a function of 
$V$ at $T=300$ K 
are presented for the FCC phase of the Ar crystal. For comparison, the calculated results 
of Schwerdtfeger \textit{et al.} \cite{PhysRevB.95.214116} as well
as the experimental results \cite{FingerStructure,Errandonea2006Structural,ross1986equation} are also included. To stay away from the melting point
(the melting pressure $P_{melt}$ and volume $V_{melt}$ at 300 K are estimated to be $1.353$ GPa  and $32.64 {\AA}^{3}$ in Refs.~\cite{WiebkeMelting,PhysRevB.65.052102}),
only the compressed volume regime with $V< 32 {\AA}^{3}$ was considered in our investigation.  Figure~\ref{Fig:P-V_diagram} shows that
our calculated $P$-$V$ results are in very good agreement with the experimental data of Ross \textit{et al.} \cite{ross1986equation} extracted
from static pressure measurements
and of Errandonea \textit{et al.} \cite{Errandonea2006Structural}
extracted from dispersive x-ray diffraction measurements up to 50 GPa. 
In addition to the FCC phase, we also calculated the $P$-$V$ curve in the high-pressure regime for the HCP phase, but its difference from that
of the FCC phase in the scale of Fig.~\ref{Fig:P-V_diagram} is tiny. Overall, it seems that our approach describes the
equation of states in high-pressure regime rather well, in contrast with the CC-based many-body expansions. Schwerdtfeger \textit{et al.} \cite{PhysRevB.95.214116} attributed the disparity of their results from experiment to the missing 4-body and higher body effects
within their approach and the possible inaccuracy of CCSD(T). In our case, RPA+rSE is less accurate than CCSD(T), but our calculations account for the many-body
effects to infinite order. From this perspective, we propose that it is most likely not the method of CCSD(T) itself, but rather the limited order
 of the many-body expansion that is responsible for the discrepancy observed by Schwerdtfeger \textit{et al.} \cite{PhysRevB.95.214116}.


\begin{figure}[t]
\centering
\includegraphics[width=0.45\textwidth,clip]{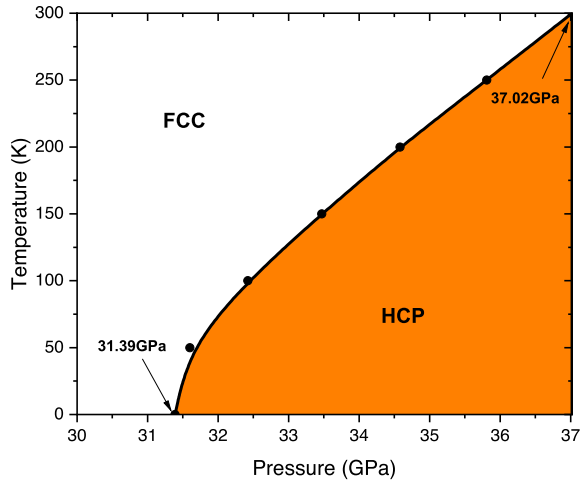}
\caption{\label{Fig:T-P_diagram} The $T-P$ diagram of the Ar crystal, where the phase boundary between the FCC and HCP phases are determined by the
Gibbs free energy. Calculations are done using the 6-atom supercell model at several successive temperatures, i.e.,  $0$, 50, 100, $\cdots$, $300$ K . 
Both electronic (RPA+rSE) and phonon (PBE) parts of the Gibbs free energy are extrapolated to the CBS limit [CBS(3,4)].}
\end{figure}

\textbf{The $T$-$P$ phase diagram for Ar.}
Figure~\ref{Fig:fcc-hcp_deltaE} signifies that at zero temperature the FCC phase can transform into the HCP phase at a compressed volume of approximately $17$ \AA$^3$ per atom, 
corresponding to a pressure of 31.4 GPa, as can be inferred from Fig.~\ref{Fig:P-V_diagram}. To investigate how the transition pressure changes with the temperature,
we calculate the Gibbs free energies $G(T,P)$ for both phases at several different temperatures, and pinpoint the boundary between the two phases in the $T$-$P$ phase diagram
 by equating the Gibbs free energies of the two phases $G_\text{FCC}(T,P)=G_\text{HCP}(T,P)$.  The resultant $T$-$P$ phase diagram is presented in
Fig.~\ref{Fig:T-P_diagram}, where the boundary line is determined by data points at seven different temperatures ranging from 0 to 300 K. One can see that the transition
pressure increases steadily from 31.4 GPa at 0 K to about 37.0 GPa at 300 K. In Table~SI of SM, we present the actual values of the critical pressure and volume along
the phase boundary line. The steady increase of the transition pressure along with the temperature is related to the different phonon spectra of the two phases. 
The lattice vibrations are such that the higher-symmetry phase (FCC) is favored over the lower-symmetry phase at finite temperatures, and thus a higher pressure is
needed to turn the FCC phase into the HCP one, compared to the zero temperature case. Close inspection reveals that the positive slope of the phase boundary is related to
the slightly larger entropy and volume of the FCC phase than the HCP phase under the same temperature and pressure conditions. Namely, the positive sign of $\Delta S/\Delta V$
across the phase boundary line gives rise to a positive $dP/dT$ slope, according to the Clausius-Clapeyon relationship (i.e., $dP/dT = \Delta S/\Delta V$) for a solid-solid
phase transition. 

Experimentally, Errandonea \textit{et al.} observed that a phase transition from FCC to HCP starts at 49.6 GPa under room temperature \cite{Errandonea2006Structural}, 
and after that the
two phases coexist over a wide pressure range. Determining the hysteresis region between the two solid-state phases requires a rather long-time $NPT$ molecular dynamics
simulations and goes beyond the scope of the present work.  Quantitatively, our simulation seems to underestimate the transition pressure, but
nevertheless predicts the correct order of magnitude for the transition pressure. Several factors might contribute to this underestimation. First, in the present work, 
the ideal structure is used for both the FCC and HCP phases, whereas in reality disorder and defects play important roles. Second, the dynamics of the phase transition
process is rather complex, governed by a delicate interplay between energetics and kinetics, as, e.g., discussed in Refs.~\cite{PhysRevLett.86.4552,PhysRevLett.96.035504} for
the martensitic FCC-HCP transition in solid xenon. These aspects have not been taken into account in our studies, but needs to be taken into account to
achieve a quantitatively accurate description of the detailed phase transition behavior.

\section{Conclusion}
To summarize, we carried out a first-principles study of the energy difference and the transition between the FCC and HCP phases of the Ar crystal. 
The electronic part of the free energy is calculated at the level of RPA plus renormalized singles corrections \cite{Ren/etal:2013}, and the ZPE and thermodynamic lattice vibrations are determined at the PBE level using the harmonic
approximation. Our results show that, at zero temperature, RPA+rSE predicts that the FCC structure has a lower energy than the HCP structure at ambient 
and high pressures up to 30 GPa. The ZPE slightly favors the FCC structure over the HCP one, but its effect is rather tiny and one order of magnitude smaller 
than the electron correlation effect described by RPA+rSE. Computationally, in order to capture the delicate energy difference between the two phases, it is important to
treat the FCC and HCP structure on an equal footing, i.e., adopting the same computational unit cell and parameters for both structures.  

In the high-pressure regime, our $P$-$V$ curve is in excellent agreement with available experimental data. Such an agreement is not achieved with the CC-based
many-body expansion approach, and has been regarded as a significant challenge for theoretical methods \cite{PhysRevB.95.214116}. Based on our results, we propose
that the inadequacy of the many-body expansion approach is probably due to the missing four- and higher-body body terms, and not due to the CCSD(T) method itself.
By computing the Gibbs free energies for both phases, we are able to determine a $T$-$P$ phase diagram of the Ar crystal. Although the transition pressure is somewhat
underestimated and no details of the transition process is provided yet, determining a qualitatively correct phase diagram for the Ar crystal
entirely from first principles itself is a great success. We believe our findings presented in this work will have significant implications for investigating 
other rare-gas systems and more complex materials in general.

\section{\label{sec:method}Methods}

\textbf{Computer Code and Computational Details.}
The ground-state total energy in this work is calculated using the RPA+rSE method,  as is implemented within the all-electron full-potential
code package FHI-aims \cite{Blum/etal:2009,Ren/etal:2012}, based on the numerical atom-centered orbital (NAO) basis set framework. 
Extensive reviews of the RPA method for many-electron ground-state energy calculations
exist in the literature \cite{Hesselmann/Goerling:2011,Eshuis/Bates/Furche:2012,Ren/etal:2012b} and its rSE correction has been discussed in
Refs.~\cite{Ren/etal:2013}. The numerical details and  benchmark tests for NAO-based RPA (and rSE) implementation for finite systems (molecules and clusters) have been presented in Refs.~\cite{Ren/etal:2012,Ren/etal:2013,Ihrig/etal:2015}. Recently,
this implementation has recently been extended to periodic systems. While the details will be presented elsewhere,
the basic numerical techniques of periodic RPA implementation within the NAO framework follow closely those of periodic $G_0W_0$ implementation published recently \cite{Ren/etal:2021}. Also, the reliability of our NAO-based periodic RPA implementation has been demonstrated by 
recently published benchmark-quality  calculations for a set of semiconductors \cite{IgorZhang/etal:2019}.

The valence correlation-consistent (VCC) NAO basis sets \cite{IgorZhang/etal:2013} are used in our calculations. Such basis sets allow one
extrapolate the calculated RPA+rSE results to the CBS limit, if the following asymptotic behavior is assumed \cite{helgaker1997basis-set}
\begin{equation}
    E(n) = E(\infty) - C/n^3
    \label{eq:basis_dependence}
\end{equation}
In Eq.~\ref{eq:basis_dependence} $n$ is the cardinal number of the NAO-VCC-$n$Z basis sets  \cite{IgorZhang/etal:2013}, and $C$ is a fitting parameter. Our CBS results 
for RPA+rSE are then obtained from a two-point extrapolation with $n=3,4$  \cite{IgorZhang/etal:2019,helgaker1997basis-set}
 \begin{equation}
  E(\infty)=\frac{E(3)3^{3}-E(4)4^{3}}{3^{3}-4^{3} }
  \label{Eq:extrapolate}
 \end{equation}
and hence denoted as CBS(3,4). The semi-local or hybrid functionals utilize only occupied state information and the convergence with respect to
the basis set is much less demanding. Hence, in the present work, the NAO-VCC-$n$Z basis set is used for these functionals. Figure.~S3 in the 
SM demonstrates the basis set convergence behavior of both RPA+rSE and PBE calculations.

As already mentioned in Sec.~\ref{sec:results}, to obtain reliable energy differences between the FCC and HCP structures, it is crucial to 
treat both structures on an equal footing. To this end,  $1\times 1\times 6$ computational cells with $ABCABC$ and $ABABAB$ stacking are used
for the FCC and HCP structures, respectively. In Fig.~S4, we show that if primitive unit cell are used instead, one may easily get in incorrect
energy ordering for the two structures for the RPA+rSE method.

The computational cost of the RPA calculations scales as $N_b^4$ with respect to the number of basis functions $N_b$. 
To reduce the computational load, the NAO-VCC-4Z RPA+rSE results for
the 6-atom computational cell at a given volume $V$ (denoted as $E_{4Z}(V)$) are obtained from the NAO-VCC-3Z results ($E_{3Z}(V)$) 
upon adding a correction term that is obtained from calculations based on the primitive unit cells,
 \begin{equation}
  E_{4Z}(V) = E_{3Z}(V) + E_{corr}(V)\, .
  \label{Eq:correction}
 \end{equation}
 Here,
  \begin{equation}
  E_{corr}(V)=6*[E_{4Z}^{prim}(V)- E_{3Z}^{prim}(V)]
  \label{Eq:correction1}
 \end{equation}
 where $E_{4Z}^{prim}(V)$ and $E_{3Z}^{prim}(V)$ are respectively the NAO-VCC-4Z and NAO-VCC-3Z results for primitive unit cells.
 In these calculations,
 $12\times 12 \times 2$, $12\times 12 \times 12$ ($12\times 12 \times 6$) $\bfk$ grids are used for the supercell and FCC (HCP) primitive cells, respectively.
 Test calculations show that such a basis correction scheme only introduces an error less than 0.1 $\mu$Ha for the energy differences between FCC
 and HCP structures, compared to direct NAO-VCC-4Z calculations for the supercell.

\textbf{ZPE and Free Energy Calculations.}
The Helmholtz free energy at a given volume $V$ and temperature $T$ invoked to determine the $P$-$V$ curve presented in Fig.~\ref{Fig:P-V_diagram} 
is given as follows,
\begin{align}
  F(V,T)=& E^\text{RPA+rSE}(V)+ \frac{1}{2}\sum_{\bfq,j}\hbar\omega_{\bfq,j}(V) + \nonumber \\
         & \emph{k}_{B}T\sum_{q,j}ln\left[2 sinh \left( \frac{\hbar\omega_{\bfq,j}(V)}{2\emph{k}_{B}T}\right) \right]
  \label{Eq:QHA}
 \end{align}
 where $E^\text{RPA+rSE}(V)$ is the RPA+rSE total energy at 0 K, and the second and third terms correspond to the contributions of ZPE
 and lattice vibrations at finite temperatures, respectively. In Eq.~\ref{Eq:QHA}, $\omega_{\bfq,j}(V)$ are the phonon frequencies,
 which in the present work are calculated using the finite difference method via the PHONOPY code \cite{phonopy} interfaced with FHI-aims.  
 The phonon spectra used to determine the ZPE and/or lattice vibration contributions
 presented in Figs.~\ref{Fig:fcc-hcp_deltaE}-\ref{Fig:T-P_diagram} are evaluated 
 using the PBE functional with a $2\times 2 \times 2 $ supercell (in units of the $1\times 1 \times 6$ computational cell, hence 48 atoms in total).  
 In the photon calculations, $12\times 12 \times 6$ $\bfk$ grid and a displacement of 0.002 \AA  (for the finite displacement method) are
 adopted. The influence of different functionals on the phonon spectra, and hence on the final ZPE and phonon contributions
 to the Helmholtz free energy is presented in Fig.~S5 of the SM.
 
 The Gibbs free energy used to determine the  $T$-$P$ phase diagram reported in Fig.~\ref{Fig:P-V_diagram} is given by
 \begin{equation}
     G(T,P) = F(V,T) + PV\, .
 \end{equation}
The Gibbs free energies are calculated separately for the FCC and HCP phases, and the phase boundary separating the two phases are
determined by equating $G_\text{FCC}(T,P)$ and $G_\text{HCP}(T,P)$ at several different temperatures.

\begin{acknowledgments}
    The work is supported by National Natural Science Foundation of China (Grant No. 11874335),  and Max Planck Partner Group for
    \textit{Advanced Electronic Structure Methods}. We thank Igor Ying Zhang and Xinzheng Li for helpful discussions.
\end{acknowledgments}

\begin{appendix}
\renewcommand{\thefigure}{S\arabic{figure}}
\setcounter{figure}{0}

\renewcommand{\thetable}{S\Roman{table}}
\setcounter{table}{0}

\newpage
\section*{Supporting Material for \\
	``Phase Stability of the Argon Crystal: A First-Principles Study Based on Random Phase Approximation plus Renormalized Single Excitation Corrections"}

\begin{figure}[h]
\centering
\includegraphics[width=0.45\textwidth,clip]{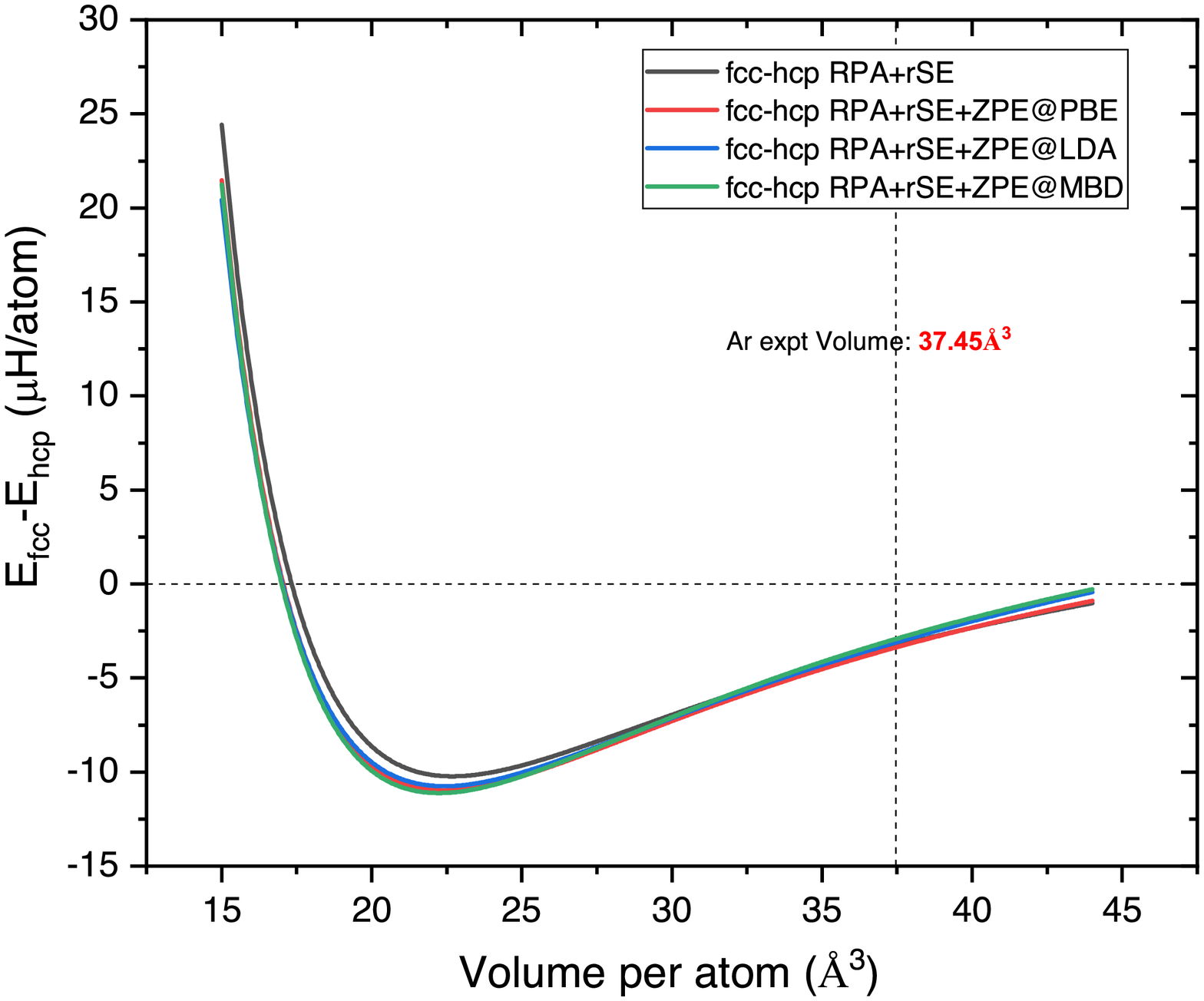}
\caption{\label{Fig:ZPE_diff_functionals} 
Energy differences between the FCC and HCP structures of the Ar crystal at zero temperature as a function of the volume (per atom). Results are obtained by  (RPA+rSE)@PBE, complemented with ZPE determined using LDA, PBE, and PBE+MBD functionals. The presented curves are obtained by subtracting the corresponding cohesive energy curves of the FCC and HCP structures,
which themselves are obtained by a second-order Birch-Murnaghan fitting of the original computed data.}
\end{figure} 
Figure~\ref{Fig:ZPE_diff_functionals} shows the energy differences between the FCC and HCP structures of the Ar crystal at zero temperature as
a function of volume, calculated using the random phase approximation (RPA) plus renormalized single excitation (rSE) correction \cite{Ren/etal:2013}. 
The RPA+rSE energy differences are further complemented by zero point energy (ZPE) as determined by three different functionals -- local-density approximation (LDA), Perdew-Burke-Erzerhof (PBE) generalized
gradient approximation (GGA), and PBE with many-body dispersion (MBD) \cite{Tkatchenko/etal:2012}. It can be seen from Fig.~\ref{Fig:ZPE_diff_functionals} the ZPE contribution 
to the energy difference is rather tiny, compared to the electron energy contribution as described by RPA+rSE. This behavior does not change if
different functionals are used to evaluate the phonon frequencies. From Fig..~\ref{Fig:ZPE_diff_functionals} we infer that even if the RPA+rSE method
is used to calculate the phonon frequencies, the scenario will not change.

\begin{figure}[h]
\centering
\includegraphics[width=0.45\textwidth,clip]{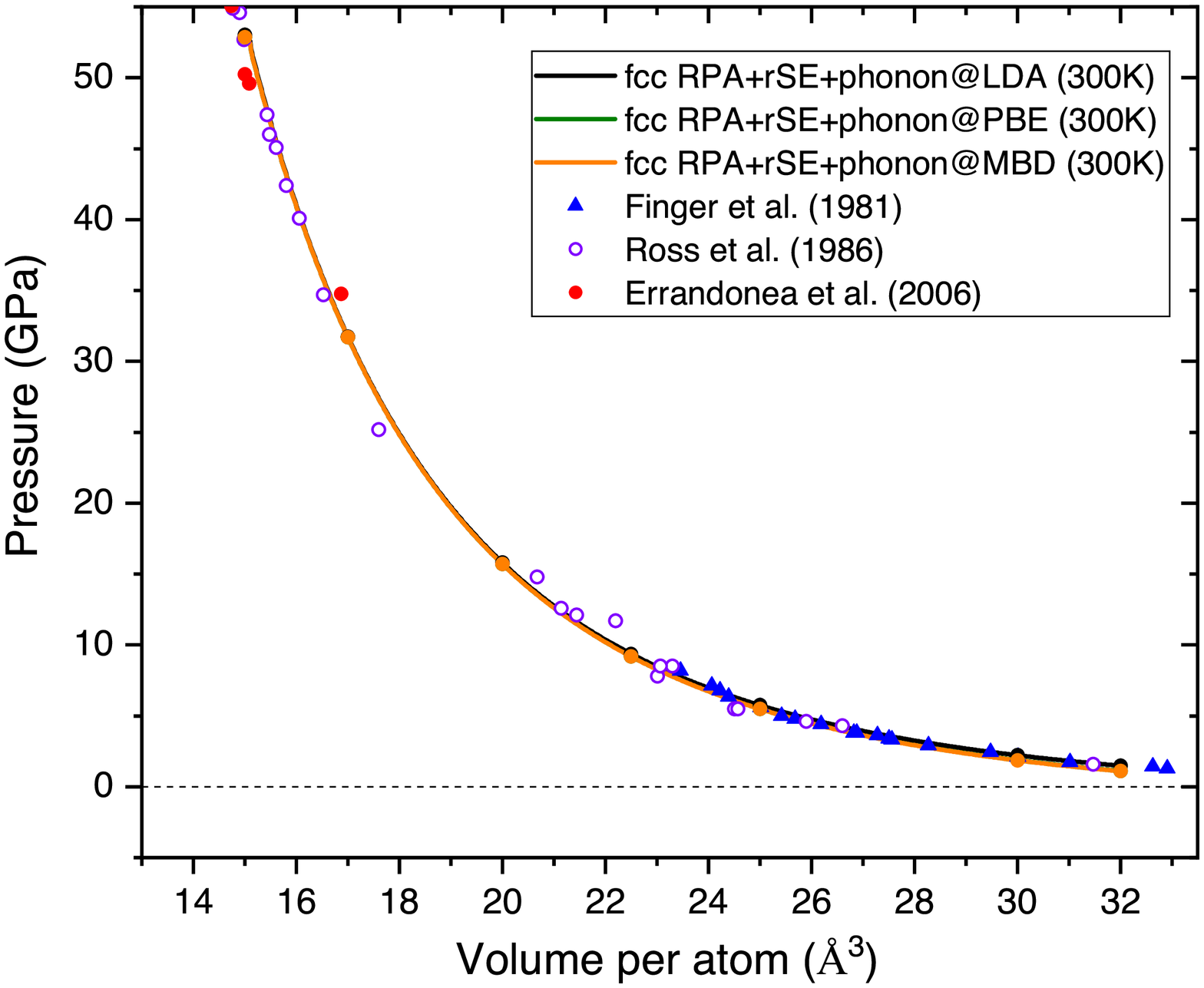}
\caption{\label{Fig:PV_diff_functionals} Calculated $P$-$V$ curves at 300 K for the FCC phase of the Ar crystal, as derived from the Helmholtz free energy. The electronic part of the Helmholtz free energy is calculated using RPA+rSE, whereas the phonon part is evaluated using
three different functionals: LDA, PBE, and PBE+MBD.  Calculations are done using the 6-atom supercell and the results are extrapolated to the complete basis set (CBS) limit. Three sets of experimental data \cite{FingerStructure,ross1986equation,Errandonea2006Structural} are also presented for comparison (cf. Fig. 3 in the main text).}
\end{figure} 

In Fig.~3 of the main text, we presented the $P$-$V$ curve of the FCC Ar crystal in the high-pressure regime. The curve is determined from the
Helmholtz free energy with respect to the volume $V$. The static (electronic) part of the free energy is calculated using (RPA+rSE)@PBE, whereas
the lattice vibration contributions are estimated from the phonon frequencies calculating the PBE functionals. Excellent agreement with the
experimental data is observed. However, there remains the question what happens if the phonon frequencies are calculated using a different
functional. In Fig.~\ref{Fig:PV_diff_functionals} we present the $P$-$V$ curves with the phonon contributions calculated using LDA, PBE,
and PBE+MBD functionals. The three curves are practically indistinguishable at the scale of Fig.~\ref{Fig:PV_diff_functionals}. This benchmark test
indicates that the obtained $P$-$V$ curve is insensitive to the underlying functionals for evaluating the phonon spectrum, and our results
are not biased because of the choice of the PBE functional.

Presented in Table~\ref{Tab:transition_pressure_PBE} are the transition pressures and volumes along the boundary line between 
the FCC and HCP phases at seven equally separated 
temperature points (0 K, 50 K, $\cdots$, 300 K). The phase boundary line is determined by equating the Gibbs free energies of the FCC and HCP phases.
The phonon contributions of the Gibbs energies are evaluated using both PBE and PBE+MBD functionals. An inspection of the results presented in the left- and right-hand sides 
of Table~\ref{Tab:transition_pressure_PBE} indicates that using different functionals to evaluate the phonon spectra has insignificant influence 
on the actual phase boundary.
\begin{table*}
\caption{Transition pressures and volumes along the phase boundary line in the $T$-$P$ phase diagram of the Ar crystal. The electronic part of the Gibbs free energy used to determine the phase boundary is calculated using (RPA+rSE)@PBE, whereas the phonon contributions are calculated using the PBE (left side) and PBE+MBD functionals (right side).}
\begin{tabular}{@{\extracolsep{\fill}}lcccccc}
\hline\hline\\[-1.5ex]
\multirow{2}{*}{Temperature(K)~~~} & \multicolumn{2}{c}{RPA+rSE+phonon@PBE}  & &
 \multicolumn{2}{c}{RPA+rSE+phonon@MBD}  \\[0.2ex]
 \cline{2-3} \cline{5-6}  \\[-1.0ex]
& ~~~~~$P_{t}$(GPa)~~~~~ & ~~~~~~$V_{t}$(${\AA}^{3}$)~~~~~ & ~~~ & ~~~~~$P_{t}$(GPa)~~~~~ & ~~~~~~$V_{t}$(${\AA}^{3}$)~~~~~   \\[0.5ex]
\hline \\[-0.5ex]
0       & 31.39   & 16.96 & & 31.79   & 16.91\\
50      & 31.60   & 16.93 & & 31.94   & 16.89\\
100     & 32.42   & 16.84 & & 32.85   & 16.79\\
150     & 33.47   & 16.73 & & 33.83   & 16.68\\
200     & 34.58   & 16.62 & & 35.03   & 16.57\\
250     & 35.81   & 16.50 & & 36.28   & 16.45\\
300     & 37.02   & 16.39 & & 37.58   & 16.33\\
\hline \\[-1.5ex]

\end{tabular}
\label{Tab:transition_pressure_PBE}
\end{table*}

\begin{figure}[h]
\centering
\includegraphics[width=0.45\textwidth,clip]{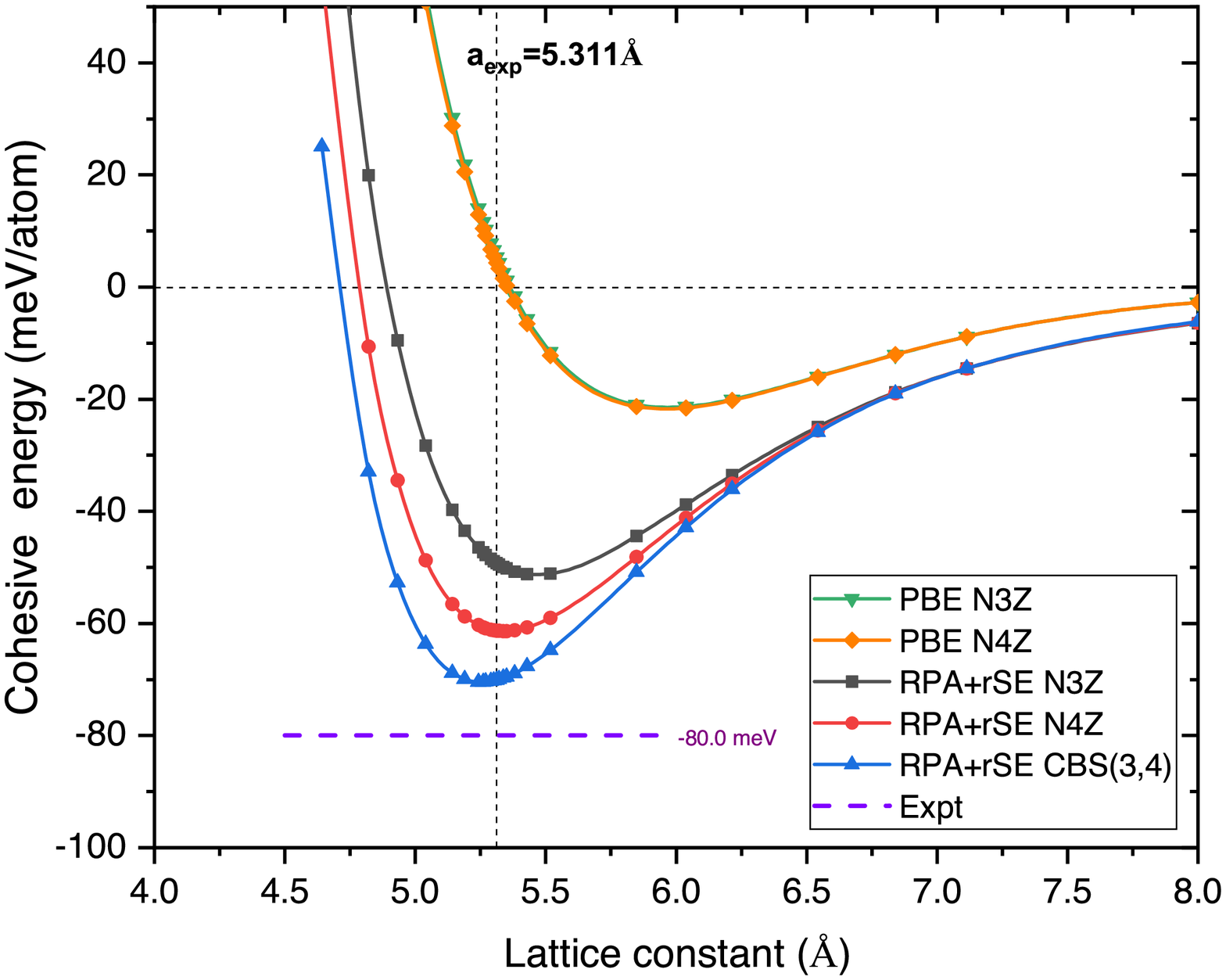}
\caption{\label{Fig:Coh_E_convergence} Calculated PBE and (RPA+rSE)@PBE cohesive energy curves at 0 K for the FCC phase of the Ar crystal, obtained
using NAO-VCC-3Z and NAO-VCC-4Z basis sets. For (RPA+rSE)@PBE, the two-point extrapolated results (CBS(3,4)) are also plotted.}
\end{figure}

In Fig.~\ref{Fig:Coh_E_convergence} we present the (RPA+rSE)@PBE and PBE cohesive energies as a function of the volume as calculated with different NAO-VCC-$n$Z \cite{IgorZhang/etal:2013} 
basis sets.
One can see that achieving basis set convergence for the correlated method (in the present case RPA+rSE) is much more demanding than the PBE functional which requires
only occupied-state information. Whereas the PBE results calculated using NAO-VCC-3Z (N3Z) and NAO-VCC-4Z (N4Z) are almost on top of each other, 
the RPA+rSE results shows a sizable shift towards stronger bonding when going from N3Z to N4Z. 
Hence, in the present work, we extrapolate the RPA+rSE results to the complete basis set (CBS) limit based on the N3Z and N4Z
results, where for PBE the N4Z results are used.

\begin{figure*}
\subfigure[Primitive unit cell]{
\includegraphics[width=0.45\textwidth]{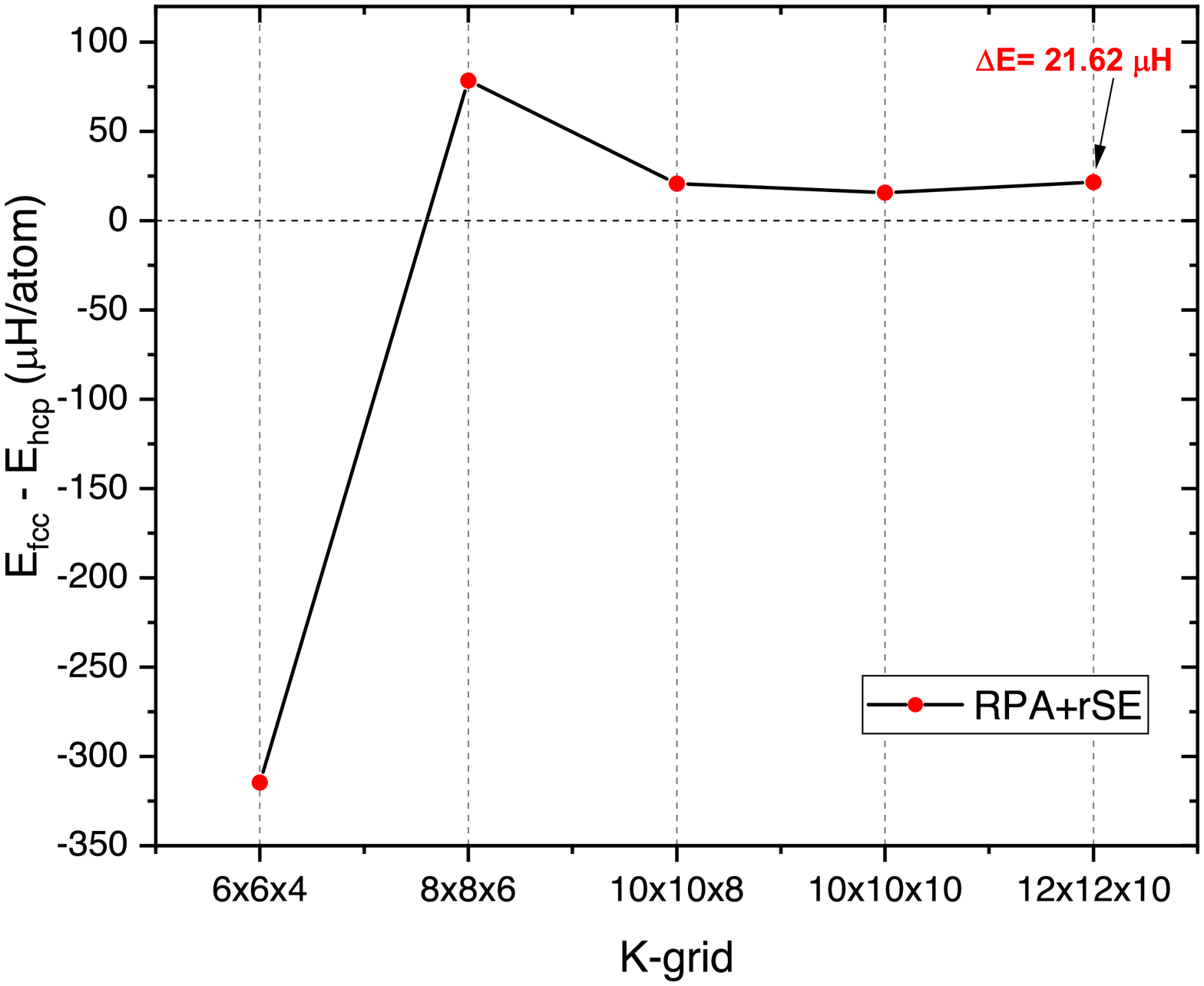}
}
\subfigure[Computational $1\times 1 \times 6$ supercell]{
\includegraphics[width=0.45\textwidth]{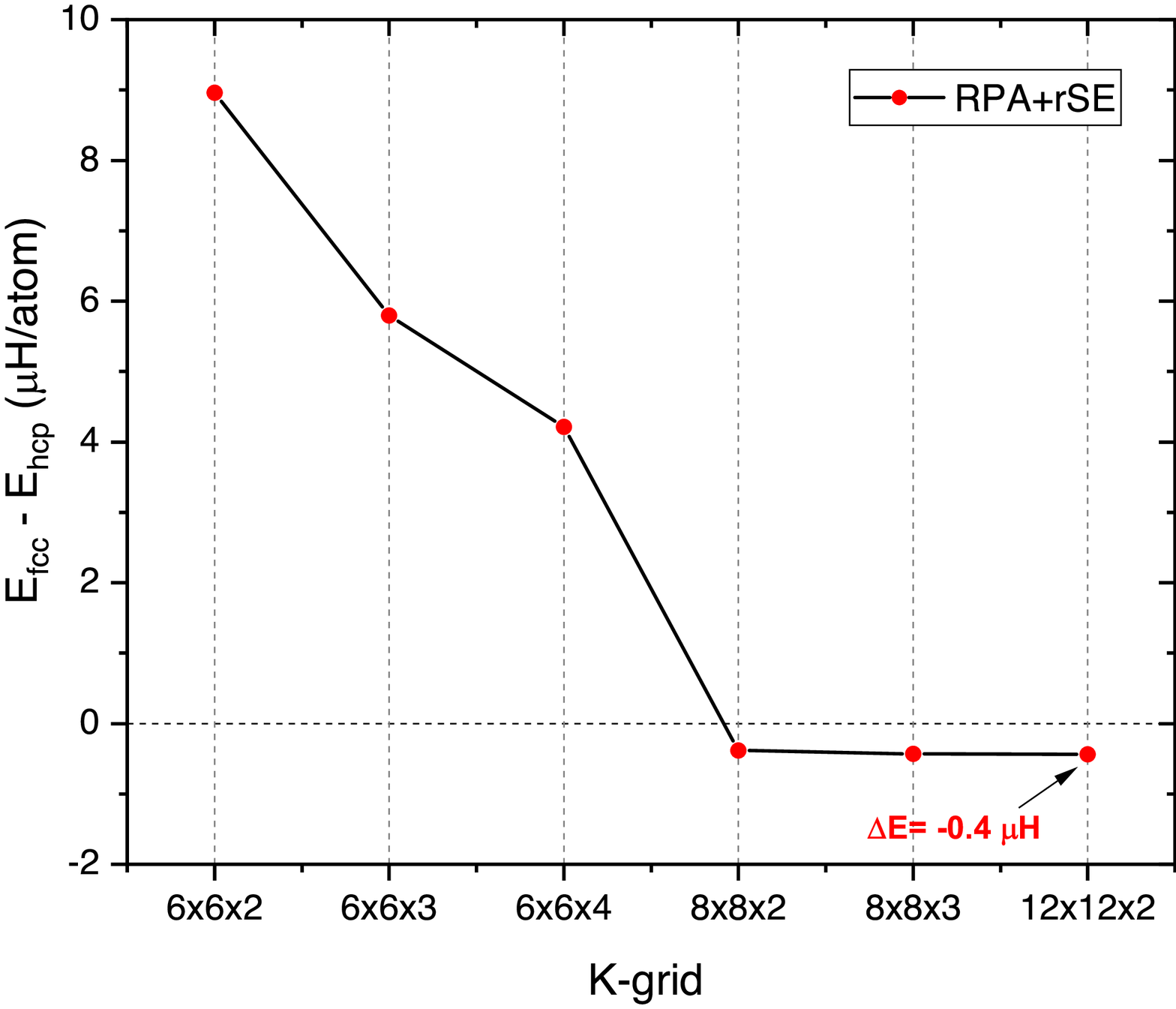}
}
\caption{\label{Fig:E_diff} Energy differences of the FCC and HCP structures with increasing $\bfk$-grid mesh: (a) Energy difference based on the primitive cell calculations; (b) energy differences based on 6-atom supercells. An atomic volume  of $V=38{\AA}^{3}$ (close to the equilibrium volume) and the NAO-VCC-3Z basis set
are used in both types of calculations.}
\end{figure*}

Figure~\ref{Fig:E_diff} compares two different strategies to compute the energy differences 
between the FCC and HCP crystal structures. 
In the first strategy, the energy differences are calculated using the respective primitive unit cells for both structures. 
In this case, as panel (a) shows, after converging $\Delta E = E_\text{fcc}-E_\text{hcp}$  with respect to the $\bfk$-grid mesh, one obtains
a positive value of $\Delta E = 21.6~\mu$Ha/atom $\approx$ 0.6 meV/atom, favoring the HCP structure. However, this is an artifact, since
HCP has a larger primitive unit cell than FCC and using the same $\bfk$ grid mesh for both structures may easily produce biased results,
especially when highly accurate results are hunted for. In the second strategy, $1\times 1 \times 6$ computational super cells (with $ABCABC$ stacking for FCC and $ABABAB$ stacking for HCP) are
used for both structures. Now by converging the energy differences with respect to the $\bfk$-grid mesh, one obtains
a $\Delta E = - 0.4~\mu$Ha/atom, marginally favoring the FCC structure. Please note this is the result obtained using an under-converged
NAO-VCC-3Z basis set. After extrapolating the results to the CBS limit, one ends up with a $\Delta E \approx - 3.5  \mu$Ha/atom, in favor
of the FCC structure.

\begin{figure}[h]
\centering
\includegraphics[width=0.45\textwidth,clip]{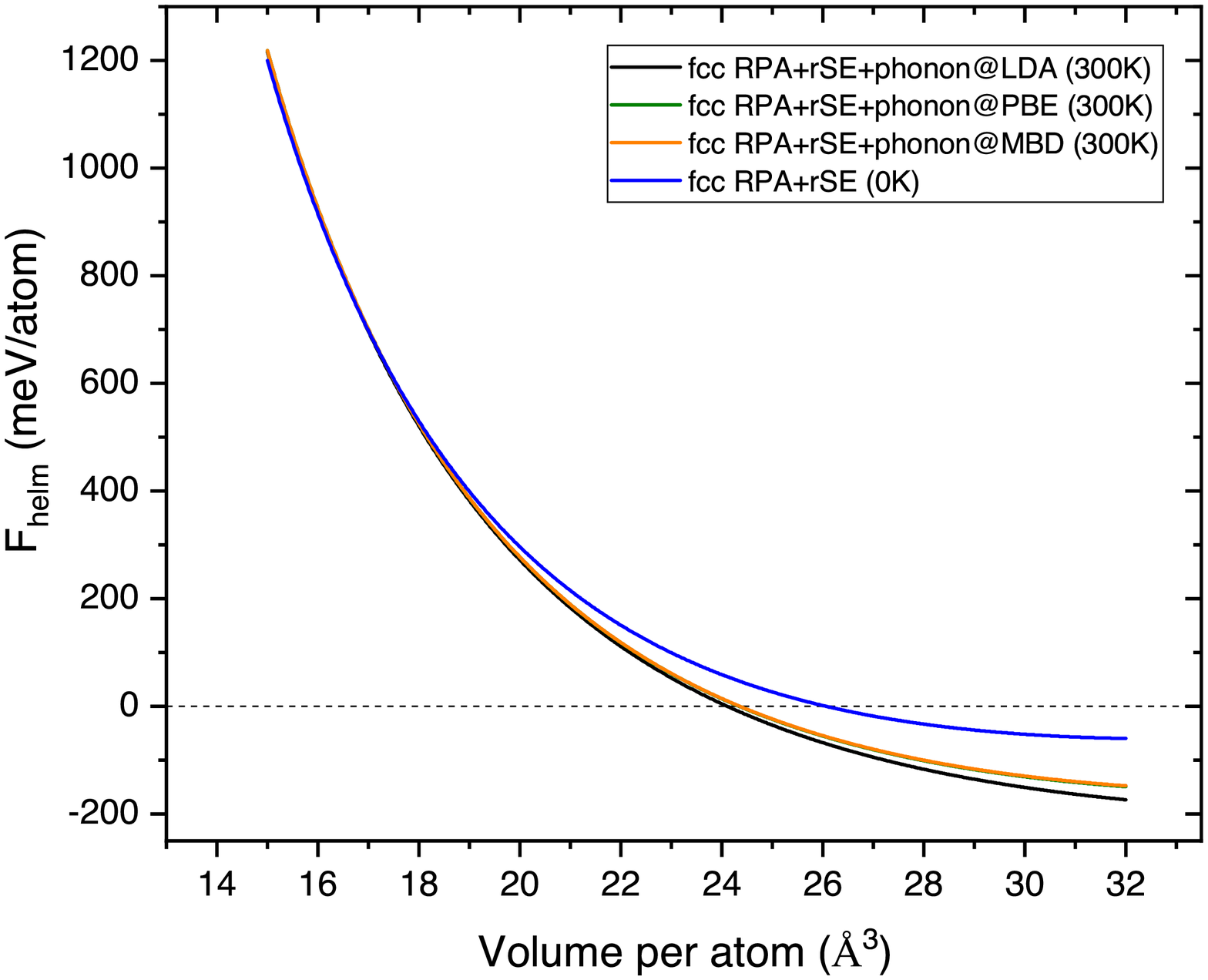}
\caption{\label{Fig:Helmholtz} Calculated Helmholtz free energy of FCC Ar crystal. Results from isolated RPA+rSE at 0K and RPA+rSE with phonon properties 
from different functionals (LDA, PBE, MBD) are also shown.}
\end{figure}
In Fig.~\ref{Fig:Helmholtz} we present the Helmholtz free energies (cf. Eq.~5 of the main text) as a function of the volume calculated at 300 K. 
Again the electronic part of the Helmholtz free
energies are calculated at the level of (RPA+rSE)@PBE, whereas the phonon contributions are evaluated using LDA, PBE, and PBE+MBD functionals. Note that
the (negative) slope of these Helmholtz free energy yields gives the $P$-$V$ curves presented in Fig.~\ref{Fig:PV_diff_functionals}.
For comparison, the
pure electronic (RPA+rSE)@PBE total energy is also plotted.  Figure~\ref{Fig:Helmholtz} reveals that the phonon contributions mainly play a role in the 
low pressure (large volume) regime. In the high pressure regime, the electronic part of the Helmholtz free energy dominates,
and it is important to adopt the RPA+rSE method in order to achieve a quantitative agreement with the experimental measurements of the equation of states (cf. Fig.~\ref{Fig:PV_diff_functionals}).
In the low pressure regime, the LDA phonons yield a noticeable difference to the Helmholtz free energy compared to the PBE phonons, whereas the MBD 
correction brings negligible changes. However, even the small differences between the LDA and PBE phonon contributions become diminishing when one
looks at the derivatives of these curves, i.e., the $P$-$V$ curves as shown in Fig.~\ref{Fig:PV_diff_functionals}.
\end{appendix}

\bibliography{CommonBib}
\end{document}